\documentstyle[emulateapj,apjfonts,psfig]{article}

\lefthead{J.~M.~Miller et al.}  \righthead{V4641~Sgr}

\received{Date}
\revised{Date}
\accepted{Date}

\journalid{vol}{date}
\articleid{1}{4}
\paperid{id}

\cpright{AAS}{1999}
\ccc{x}

\begin{document}

\title{A Relativistic Fe~K$\alpha$ Emission Line in the
Intermediate Luminosity \textit{BeppoSAX} Spectrum of the Galactic
Microquasar V4641~Sgr}

\author{J.~M.~Miller\altaffilmark{1}, 
        A.~C.~Fabian\altaffilmark{2},
	J.~J.~M.~in~'t~Zand\altaffilmark{3,4},
	C.~S.~Reynolds\altaffilmark{5},
	R. Wijnands\altaffilmark{1,6},
	M.~A.~Nowak\altaffilmark{1},
	and W.~H.~G.~Lewin\altaffilmark{1}
	}

\altaffiltext{1}{Center~for~Space~Research and Department~of~Physics,
        Massachusetts~Institute~of~Technology, Cambridge, MA
        02139--4307; jmm@space.mit.edu}
\altaffiltext{2}{Institute of Astronomy, University of Cambridge,
        Madingley Road, Cambridge CB3 OHA, UK}
\altaffiltext{3}{Astronomical Institute, Utrecht University, P.O. Box 80000, NL -- 3508 TA Utrecht, the Netherlands}
\altaffiltext{4}{SRON National Institute for Space Research, Sorbonnelaan 2, NL -- 3584 CA Utrecht, the Netherlands}
\altaffiltext{5}{Department of Astronomy, University of Maryland,
        College Park, MD 20742}
\altaffiltext{6}{Chandra Fellow}


\authoremail{jmm@space.mit.edu}

\label{firstpage}

\begin{abstract}
Broad Fe~K$\alpha$ emission lines have recently been reported in a
number of Galactic black holes.  Such lines are useful accretion flow
diagnostics because they may be produced at the inner accretion disk
and shaped by relativistic effects, but in general they have only been
observed at luminosities of $L_{X} \sim 10^{37-38}~{\rm erg}~{\rm
s}^{-1}$ in soft X-rays.  The Galactic microquasar V4641~Sgr ---
widely known for its 12.2~Crab (1.5--12 keV) outburst in 1999
September --- displayed low-level activity in 1999 March.
\textit{BeppoSAX} observed the source in this state and Fe~K$\alpha$
line emission was found (in~'t Zand et al. 2000).  In re-analyzing
these data, we find strong evidence that the Fe~K$\alpha$ line profile
is broadened.  For the most likely values of the source distance and
black hole mass measured by Orosz et al. (2001), our fits to the total
spectrum indicate that the source was observed at a luminosity of
$L_{X} = 1.9^{+1.0}_{-0.8} \times 10^{36}~{\rm erg}~{\rm s}^{-1}$
(2--10 keV), or $L_{X}/L_{Eddington} = 1.8^{+0.9}_{-0.8} \times
10^{-3}$.  Advection-dominated accretion flow (ADAF) models predict a
radially-recessed disk in this regime.  In contrast, fits to the
observed Fe~K$\alpha$ emission line profile with a relativistic line
model constrain the inner disk to be consistent with the marginally
stable circular orbit of a Schwarzschild black hole.
\end{abstract}

\section{Introduction}
Recently, studies of the Fe~K$\alpha$ line profiles in Galactic black
holes and black hole candidates have revealed strong, skewed lines,
similar to those that are observed in some Seyfert galaxies (see,
e.g., Fabian et al. 2000; Weaver, Gelbord, \& Yaqoob 2001).  Such
lines are likely produced by the irradiation of an accretion disk
which extends close to the marginally stable orbit around the black
hole.  The exact shape of such lines can indicate the inner extent of
the disk (and thereby constrain black hole spin), its ionization
state, and the nature of the interaction between the disk and the
corona that irradiates its surface.

The broad Fe~K$\alpha$ lines recently reported have all been observed
in bright persistent systems (Cygnus~X-1: Miller et al. 2002a,
GRS~1915$+$105: Martocchia et al. 2002) or in bright states of
transient systems (XTE~J1650$-$500: Miller et al. 2002b,
SAX~J1711.6$-$3808: in~'t~Zand et al. 2002).  These lines provide
important new constraints on the black hole spin and inner accretion
flows of these systems.  While most models for the accretion flow
geometry in Galactic black holes predict that the disk extends near to
the marginally stable orbit in bright states ($L_{X}\simeq
10^{37-38}$~erg/s, or higher), the nature of lower-luminosity states
may be different.  In particular, ADAF models for transient systems
(e.g., Esin, McClintock, \& Narayan 1997) predict that the disk may be
recessed to several tens or hundreds of gravitational radii ($R_{g} =
G M_{BH} / c^{2}$).

V4641~Sgr is usually identified with its remarkable 12.2~Crab outburst
lasting only a few hours on 1999 September 15 (Smith et al. 1999).
Dynamical constraints strongly indicate a black hole primary ($5.49
\leq M_{BH} \leq 8.14~M_{\odot}$), a high inclination ($60^{\circ}
\leq i \leq 71^{\circ}$), and a large distance ($7.40\leq d\leq
12.31$~kpc, Orosz et al. 2001).  Hjellming et al. (2000) monitored the
outburst in radio and found jets expanding at apparently superluminal
velocities; V4641~Sgr is therefore termed a ``microquasar.''
Significant low-luminosity activity was observed by \textit{BeppoSAX}
and by \textit{RXTE} Galactic bulge scans (Swank \& Markwardt 2001)
earlier in 1999, however, well before the better-known outburst.
In~'t Zand et al. (2000) reported on a pointed \textit{BeppoSAX}
observation made on 1999 March 13.  A line complex was noted, and
attributed to Fe~XXV and Fe~XXVI lines with widths narrower than the
\textit{BeppoSAX} resolution.  The best-fit time-averaged 2--10~keV
flux measured on March 13 was found to be $2.7\times 10^{-10}~{\rm
erg}~{\rm cm}^{-2}~{\rm s}^{-1}$, corresponding to $1.8\times
10^{36}~{\rm erg}~{\rm s}^{-1} \leq L_{X} \leq 4.9\times 10^{36}~{\rm
erg}~{\rm s}^{-1}$.  Motivated by the constraints on the accretion
flow geometry made through the study of broad lines in other sources
at higher luminosities, we re-visit this intermediate-luminosity
spectrum of V4641~Sgr.

\section{Observation and Data Reduction}
\textit{BeppoSAX} observed V4641~Sgr on 1999 March 13.22.  The
soft-X-ray-imaging narrow-field instruments (NFI) aboard
\textit{BeppoSAX} are called the Low-Energy and Medium-Energy
Concentrator Spectrometer (LECS and MECS, respectively).  The net
exposure time for the LECS was 10.2~ks; the net MECS exposure was
27.7~ks.  The MECS has an energy resolution of 8\% at 6~keV.  

The Phoswich Detector System (PDS; 12--300~keV) observed V4641~Sgr
simultaneously with the LECS and MECS.  The statistical quality of the
high energy spectrum obtained is relatively poor, and does not allow
for tighter constraints on the continuum.  In particular, reflection
models --- which describe a geometry in which the Fe~K$\alpha$
emission line and a ``Compton hump'' result from the irradiation of
the disk --- are not better-constrained by considering the PDS
spectrum.  Our analysis is therefore limited to the time-averaged NFI
spectra.

All aspects of the source and background spectral extraction,
instrument response function generation, and filtering are exactly the
same as reported in in~'t~Zand et al. (2000).  We considered the LECS
spectrum between 0.4 and 4.0~keV, the MECS spectrum between 2.0 and
10.0~keV (again, similar to in~'t Zand et al. 2000).  The spectra were
fit using XSPEC version 11.1 (Arnaud 1996).  All errors reported in
this work are 90\% confidence errors.  The LECS and MECS spectra were
fit jointly.  An overall normalizing constant was allowed to float
between the spectra to account for differences in the instrumental
flux calibration.

\section{Analysis and Results}
Preliminary fits suggested that the absorbing column and the nature of
the soft component could not be constrained independently.  We
therefore fixed the equivalent neutral hydrogen column density at the
weighted average measured along this line of sight by Dickey \&
Lockman (1990): $N_{H}=2.3\times 10^{21}~{\rm atoms}~{\rm cm}^{-2}$
(assuming the ``phabs'' model in XSPEC).

We attempted to describe the continuum spectra with a number of
models, including the ``bulk motion Comptonization'' model (Shrader \&
Titarchuk 1999), a model consisting of a multicolor disk black-body
(MCD; Mitsuda et al. 1984) and power-law components, and a model
consisting of MCD and ``comptt'' (Titarchuk 1994) components (as per
in~'t~Zand et al. 2000).  None of these models gives an acceptable fit
to the data, yielding $\chi^{2}/\nu =~~6.22,~5.83,~{\rm and}~5.96$,
respectively ($\nu$ is the number of degrees of freedom; $\nu=76$ for
the first two models, $\nu=74$ for the last).  These continuum models
all fail to account for a strong emission feature in the Fe~K$\alpha$
line region.

In Figure 1, the line profile is shown as a ratio of the spectrum to
these continuum models.  Following Iwasawa et al. (1999), the ratio is
formed by ignoring the 4--7~keV range when fitting the spectra.  It is
apparent that the line may not be intrinsically narrow --- the profile
revealed in the data/model ratio is similar to those expected for a
line produced by irradiation of the inner accretion disk around a
black hole (Fabian et al. 1989, Laor 1991).  In a previous examination
of this spectrum, in~'t Zand et al. (2000) modeled the Fe~K$\alpha$
line region with two Gaussians (at 6.68~keV and 6.97~keV, as per
Fe~XXV and Fe~XXVI) --- each with zero widths --- assuming that the
intrinsic width of each line is below the instrumental resolution.
The bottom panel in Figure 1 shows the data/model ratio obtained with
this narrow line model; clearly, flux between 5.0--6.4~keV --- perhaps
associated with an intrinsically broad line --- is not accounted for
by this model.

Adopting the MCD plus power-law model for the continuum, we next
investigated the nature of the Fe~K$\alpha$ line region in detail.
For a line model consisting of two zero-width Gaussians, $\chi^{2}/\nu
= 1.721, (\nu = 74)$.  However, for fits with the Laor

\centerline{~\psfig{file=f1.epsi,width=3.5in,angle=0}~}
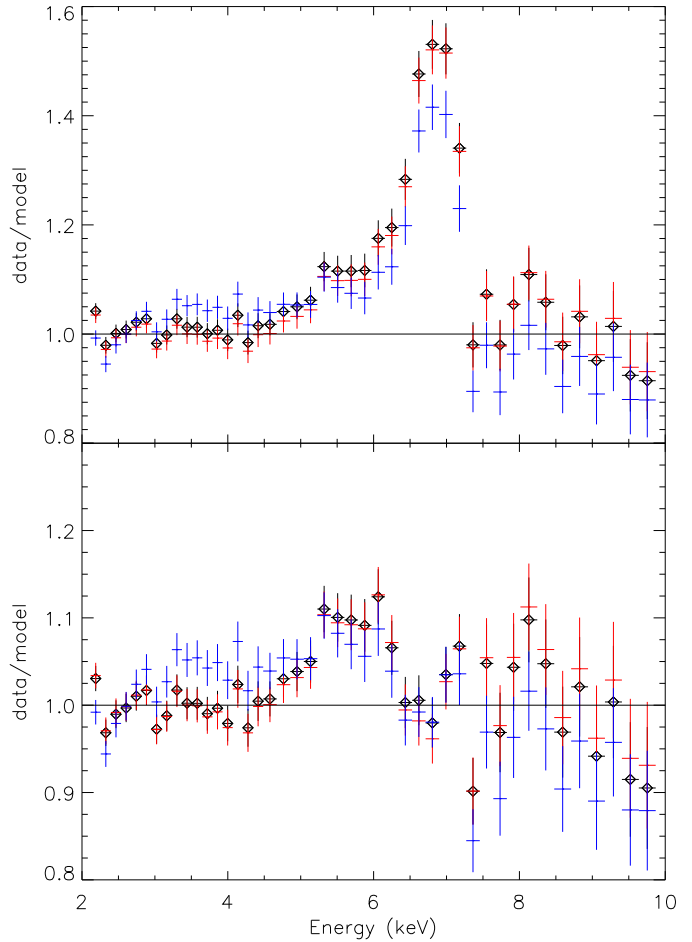
\figcaption[h]{\small Data/model ratios for the MECS spectra and a
number of viable of continuum models.  In black (and with diamonds),
the ratio for an MCD plus power-law continuum.  In red, the ratio for
an MCD plus ``comptt'' continuum.  In blue, the ratio for the ``bulk
motion Comptonization'' model.  \textit{Above}: The 4--7~keV range was
ignored in fitting the spectrum.  The red wing of the line is clearly
evident in the ratio of the data to each continuum model.
\textit{Below}: two zero-width Gaussian lines added at 6.68~keV and
6.97~keV were added to the models (corresponding to narrow Fe~XXV and
Fe~XXVI emission lines).  The red wing between $\sim$5--6.4~keV is not
accounted for by these Gaussians.}
\medskip

line model, $\chi^{2}/\nu = 1.230, (\nu = 71)$.  The improvement in
the fit statistic is significant at the 4.5$\sigma$ level of
confidence.  This continuum plus line model is our best-fit spectral
model, and it is shown in Figure 2.  The inclusion of a smeared Fe~K
edge (``smedge,'' Ebisawa et al. 1994) does not improve the fit
significantly.  When a smeared edge is included, the edge parameters
are not well-constrained: $E_{edge} = 9.1^{+0.2}_{-1.1}$~keV and $\tau
= 1.6^{+2.7}_{-1.4}$.  Allowing a standard narrow edge component to
float in the Fe~K$\alpha$ region does not give an improvement in the
fit statistic.

The Laor line energy was only permitted to vary within the range
$6.40~{\rm keV} \leq {\rm E}_{Laor} \leq 6.97~{\rm keV}$ (Fe~I --
Fe~XXVI).  The best-fit energy is measured to be ${\rm E}_{Laor} =
6.40^{+0.3}$~keV.  The line emitting region is relatively well
constrained: $R_{in} = 8^{+4}_{-3}~R_{g}$ and $ R_{out} =
30^{+40}_{-10}~R_{g}$ ($R_{g} = G M_{BH} / c^{2}$).  As is evident
from the line profile shown in Figure 1, the line equivalent width is
very high: $W_{K \alpha} = 870^{+100}_{-70}$~eV ($9.8^{+1.2}_{-1.0}
\times 10^{-12}~{\rm erg}~{\rm s}^{-1}$, 0.4--10~keV).  The
inclination is measured to be $i = 43^{\circ} \pm 15^{\circ}$.  This
inclination is lower than the optically-determined range; however,
Orosz et al. (2001) note that the inclination implied by the jet is
very 

\centerline{~\psfig{file=f2.ps,width=3.5in,angle=-90}~}
\centerline{~\psfig{file=f3.ps,width=3.5in,angle=-90}~}
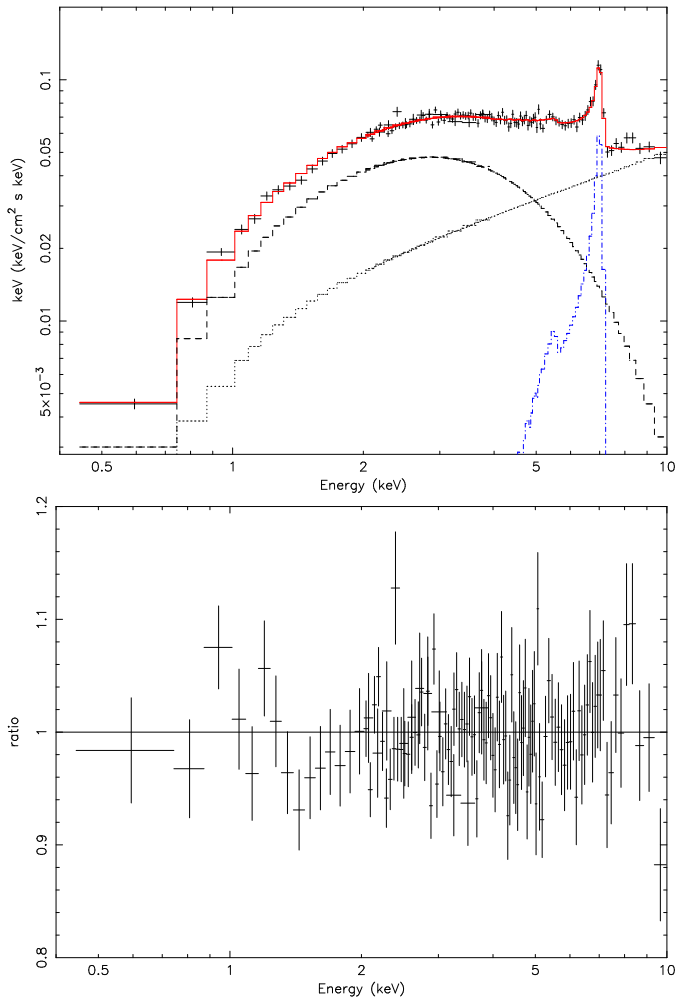
\figcaption[h]{\small \textit{Above}: The unfolded 0.4--10.0~keV NFI
spectrum and best-fit model (the spectra shown are normalized to have
the same flux).  In red, the total spectral model is shown.  The Laor
line model is shown in blue.  \textit{Below}: The data/model ratio for
best-fit model shown above.}
\medskip

low ($i \leq 6^{\circ}$).  In this context, the intermediate value we
measure is acceptable.  The inner disk emissivity profile ($J(r)\sim
r^{-\beta}$) measured with this model is consistent with that expected
for a standard accretion disk ($\beta = 3$).

With the Laor model for the Fe~K$\alpha$ line region, we find that the
MCD inner disk color temperature is: $kT = 1.14\pm 0.03$~keV with a
normalization of $N_{MCD} = 4.5\pm 0.5$; this component contributes an
unabsorbed flux of $1.5\pm 0.2 \times 10^{-10}~{\rm erg}~{\rm
cm}^{-2}~{\rm s}^{-1}$ in the 0.4--10~keV band.  The power-law index
is measured to be $\Gamma = 1.3\pm 0.1$ with a normalization of
$N_{pl} = 1.1\pm 0.3 \times 10^{-2}~ ({\rm photons}~ {\rm keV}^{-1}~
{\rm cm}^{-2}~ {\rm s}^{-1}~ {\rm at 1~keV})$; this component contributes
an unabsorbed flux of $1.1\pm 0.2 \times 10^{-10}~{\rm erg}~{\rm
cm}^{-2}~{\rm s}^{-1}$ in the 0.4--10~keV band.

For the most likely values of the source distance and black hole mass
measured by Orosz et al. (2001), we measure a total source luminosity
of $L_{2-10} = 1.9^{+1.0}_{-0.8} \times 10^{36}~{\rm erg}~{\rm
s}^{-1}$, or $L_{0.4-10} = 3.2^{+1.7}_{-1.4} \times 10^{36}~{\rm
erg}~{\rm s}^{-1}$ (0.4--10.0~keV).  As a fraction of the Eddington
luminosity, these values correspond to $L_{2-10}/L_{Edd.} =
1.8^{+0.9}_{-0.8} \times 10^{-3}$ and $L_{0.4-10}/L_{Edd.} =
3.0^{+1.6}_{-1.3} \times 10^{-3}$, or $\dot{m}_{Edd.} \gtrsim
3.0^{+1.6}_{-1.3} \times 10^{-2}$ (defining $\dot{m}_{Edd.} =
10~L_{bolometric} / c^{2}$; equivalent to assuming an efficiency of
10\%).  Extending these values to a broader range depends slightly on
the spectral model used, and the data at high energies is not very
constraining.  For the model adopted by in~'t~Zand et al. (2000) and
our best-fit model, the 0.4--120~keV luminosity is approximately
double the 0.4-10~keV luminosity.

Although fits with three continuum models commonly applied to Galactic
black hole spectra indicate a broad Fe~K$\alpha$ emission line, we
have found two sets of fits that indicate instead a narrow line.  The
first set of fits invoke a partial covering model.  For example,
modeling absorption in the ISM with ``phabs'' ($N_{H, ISM} =
4.8^{+0.4}_{-0.2} \times 10^{21}~{\rm cm}^{-2}$) and allowing two
partial covering models (``pcfabs'' in XSPEC; $N_{H, 1} = 4.5 \pm 1.5
\times 10^{22}~{\rm cm}^{-2}~{\rm and}~ f_{1} = 0.35 \pm 0.10$, $N_{H,
2} = 2.7 \pm 0.3 \times 10^{21}~{\rm cm}^{-2}~{\rm and}~ f_{2} = 0.66
\pm 0.05$) and a simple power-law continuum ($\Gamma = 2.7 \pm 0.1$)
gives an acceptable fit ($\chi^{2}/\nu = 1.10, \nu = 70$) with only a
narrow Gaussian line (E $= 6.89 \pm 0.04$~ keV, FWHM $=
0.1_{-0.1}^{+0.1}$~ keV).  Partial covering models are commonly fit to
neutron star ``accretion disk corona'' sources which display strong
dipping behavior in their X-ray lightcurves (see, e.g., Morley et
al. 1999).  Dips have not been reported in V4641~Sgr.  It is unlikely
that partial covering models are justified for this source.

The second set of fits involves the ``compps'' Comptonization model of
Poutanen \& Svensson (1996).  For the models discussed above that
utilized separate soft and hard phenomenological fit components, there
is some worry that the red tail of the broad line is an artifact of
the region where the soft and hard fit components cross.  We have used
compps to model a disk blackbody spectrum Comptonized by a slab corona
(with unity covering fraction), and reflected off of a \emph{neutral},
cold medium with Fe abundance relative to hydrogen that is three times
solar. An $\approx 40$\,keV corona with $\tau_{\rm es} \approx 0.9$
fits the continuum well, whereas the line region is well-described by
a narrow 6.9\,keV line with $\approx 250$\,eV equivalent width.
However, an extremely large reflection fraction of $R\approx 3$ is
then required.  A very large neutral Fe edge from reflection coupled
with a narrow Fe {\sc XXVI} line or K$\beta$ line from more neutral Fe
would be fairly unusual.  We note, however, that a very similar compps
model with a somewhat lower reflection fraction ($R \approx 1.5$),
does allow for a relativistic line with rest energy 6.9\,keV,
emissivity index $\approx -2.2$, equivalent width 360\,eV, and an
inner emission radius of $6\,GM/c^2$. The lack of reliable data above
10\,keV prevents us from exploring further these models requiring
large reflection.

It is difficult to identify the ``state'' in which this observation of
V4641~Sgr occurred (for a recent discussion of Galactic black hole
states, see Homan et al. 2001; for a review see Done 2002).
In~'t Zand et al. (2000) report no measurable narrow or broad-band
features in the power density spectrum (PDS) of the MECS light curve,
consistent with the ``high/soft'' state.  This state is defined by a
soft, thermal, disk-dominated spectrum, with a weak, soft ($\Gamma
\sim 3$) power-law; we measure $\Gamma = 1.3 \pm 0.1$.  The
intermediate luminosity and relatively hot accretion disk ($kT =
1.14\pm 0.03$~keV, representing 58\% of the 0.4--10~keV flux) are
inconsistent with a ``low/hard'' state identification.  The lack of
broad-band noise in the PDS and the extremely hard power-law index are
inconsistent with both the ``very high'' and ``intermediate'' states,
though the limits on the noise are not very constraining.  We suggest
that V4641~Sgr was observed in a kind of ``intermediate'' state, which
may differ from similar states in other Galactic black hole systems.

\section{Discussion}
We have re-analyzed the \textit{BeppoSAX} NFI spectra of the Galactic
microquasar V4641~Sgr obtained on 1999 March 13.  Applying the Laor
model for line emission from the inner part of an accretion disk
around a black hole, we find that a broad, relativistically-shaped
Fe~K$\alpha$ line is preferred over a sum of narrow helium-like and
hydrogenic Fe~K$\alpha$ lines at the 4.5$\sigma$ level of confidence.
A line model --- whether a complex of narrow lines or a broadened line
--- is strongly required by the data.  Within the context of the Laor
model, the narrow emission observed near 6.8~keV (see Figure 1) is the
blue wing of a relativistically-shaped line from more neutral Fe
species including Fe~I~K$\alpha$ at 6.40~keV (see Figure 2, Section
3).  The line is among the strongest yet observed in a Galactic black
hole system: $W_{K \alpha} = 870^{+100}_{-70}$~eV.  A spinning black
hole (implied if $R_{in} < 6~R_{g}$) is allowed but not required by
the line model; our best-fit value is $R_{in} = 8^{+4}_{-3}~R_{g}$.
Although the line may be broadened partially by Compton processes, it
is likely that relativistic shifts dominate as the profile is strongly
asymmetric.

In most cases, observations of Galactic black holes at $L_{X} \simeq
10^{36}~{\rm erg}~{\rm s}^{-1}$ do not achieve the sensitivity
required to clearly detect Fe~K$\alpha$ emission lines or fast
quasi-periodic oscillations ($few \times 100$~Hz; plausibly associated
with the Keplerian frequency near the marginally stable orbit).  The
Fe~K$\alpha$ emission line we have observed in V4641~Sgr, then,
represents a rare opportunity to test accretion flow models.

Esin, McClintock, \& Narayan (1997) predict that at $\dot{m}_{Edd.}
\leq 0.08$, the inner accretion flow in Galactic black holes will be
radiatively inefficient and the disk will be recessed to $R_{in} \geq
200~R_{g}$.  The line we have observed, however, is consistent with
being produced in a disk with $R_{in} = 8^{+4}_{-3}~R_{g}$.  In the
0.4--10.0~keV band, we measure $\dot{m}_{Edd.} = 3.0^{+1.6}_{-1.3}
\times 10^{-2}$ (assuming the most likely values for the black hole
mass and distance measured by Orosz et al. 2001).  This value
approximately doubles on the 0.4--120~keV range; the value of
$\dot{m}_{Edd.}$ given by a bolometric luminosity might be a factor of
a few higher.  We also note that taking extremes for the black hole
mass, distance, and observed flux can yield values of $\dot{m}_{Edd.}$
that are significantly lower than than the range we quote, and also
above the critical value of $\dot{m}_{Edd.} = 0.08$.  It is clear,
however, that in the regime where an ADAF is expected to begin to take
hold, the observed Fe~K$\alpha$ line profile strongly suggests a disk
which extends close to the marginally stable orbit.

The Fe~K$\alpha$ line we have measured is not the strongest yet
observed from V4641~Sgr.  \textit{RXTE}/PCA spectra obtained during
the decline of the 12.2~Crab flare ($L_{X} \sim 3-4 \times
10^{39}~{\rm erg}~{\rm s}^{-1}$) in 1999 September, reveal an
Fe~K$\alpha$ line at $\sim 6.70$~keV with $W_{K \alpha} \sim 2.4$~keV
(Revnivtsev et al. 2002).  The resolution of the PCA is not well-suited
to reveal a line profile of the kind we find in the 1999 March NFI
spectra.  In part, the large equivalent width we have measured may be
explained by an ionized disk (Ballantyne, Ross, \& Fabian 2002).  We
also speculate that Fe may be over-abundant (relative to solar) by a
factor of a few in this source.  Orosz et al. (2001) report that N and
Ti may be over-abundant by a factor of 10, Mg by a factor of 7, O by a
factor of 3, and Ca by a factor of 2.  

The Fe~K$\alpha$ emission line observed in an \textit{XMM-Newton}
spectrum of the Galactic black hole candidate XTE~J1650$-$500 extends
to $\sim$4~keV and implies a spinning black hole (Miller et
al. 2002b).  The line profile seen in the NFI spectra of V4641~Sgr is
not as strongly skewed (perhaps extending to $\sim$5~keV; see Figure
1) as the profile observed in XTE~J1650$-$500, likely indicating that
the line is shaped less by strong gravitational effects.  In an NFI
observation of the Galactic microquasar GRS~1915$+$105, Martocchia et
al. (2002) find a relativistically-shaped Fe~K$\alpha$ line consistent
with neutral ion species, which does not require a spinning black hole
(based on the implied inner extent of the accretion disk).
Qualitatively, the line profile observed in GRS~1915$+$105 may be the
most similar to that in V4641~Sgr.  \textit{RXTE}/PCA spectra of the
Galactic black hole candidate GX~339$-$4 at $L_{X} \sim few~\times
10^{36}~{\rm erg}~{\rm s}^{-1}$ also reveal evidence for broad,
relativistically-shaped line profiles (Nowak, Wilms, \& Dove 2002)
that render an ADAF geometry unlikely.

\section{Acknowledgments}
We thank the anonymous refreee for helpful suggestions.
R. W. was supported by NASA through Chandra fellowship grant
PF9-10010, which is operated by the Smithsonian Astrophysical
Observatory for NASA under contract NAS8-39073.
W. H. G. L. gratefully acknowledges support from NASA.  This research
has made use of the data and resources obtained through the HEASARC
on-line service, provided by NASA-GSFC.


\begin{references}

\reference{} Arnaud, K. A., 1996, Astronomical Data Analysis Software
and Systems V, eds. G. Jacoby and J. Barnes, p17, ASP Conf. Series
vol. 101

\reference{} Ballayntyne, D. R., Ross, R. R., \& Fabian, A. C., 2002,
MNRAS, 329, L67

\reference{} Dickey, \& Lockman, 1990, ARAA, 28, 215

\reference{} Done, C., 2002, in ``Philosophical Transactions of the
Royal Society,'' Series A: Mathematical, Physical, and Engineering
Sciences, astro-ph\/0203246

\reference{} Ebisawa, K., et al., 1994, PASJ, 46, 375

\reference{} Esin, A. A., McClintock, J. E., \& Narayan, R., 1997,
ApJ, 555, 489, 865

\reference{} Fabian, A. C., Rees, M. J., Stella, L., \& White, N. E.,
1989, MNARAS, 238, 729

\reference{} Fabian, A. C., Iwasawa, K., Reynolds, C. S., and Young,
A. J., 2000, PASP, 112, 1145

\reference{} Hjellming, R. M., et al., 2000, ApJ, 544, 977

\reference{} Homan, J., Wijnands, R., van der Klis, M., Belloni, T.,
van Paradijs, J., Klein-Wolt, M., Fender, R., \& Mendez, M., 2001,
ApJS, 132, 377

\reference{} Iwasawa, K., Fabian, A. C., Young, A. J., Inoue, H.,
and Matsumoto, C., 1999, MNRAS, 306L, 19 

\reference{} Laor, A., 1991, ApJ, 376, 90

\reference{} Martocchia, A., Matt, G., Karas, V., Belloni, T., \&
Feroci, M., 2002, A \& A., subm., astro-ph/0203185


\reference{} Miller, J. M., et al., 2002a, ApJ, in press,
astro-ph/0202083

\reference{} Miller, J. M., et al., 2002b, ApJ, 570, L69

\reference{} Mitsuda, K., et al., 1984, PASJ, 36, 741

\reference{} Morley, R., Church, M. J., Smale, A. P., \&
Baluckinska-Church, M., 1999, MNRAS, 302, 593

\reference{} Nowak, M. A., Wilms, J., \& Dove, J. B., 2002, MNRAS, in press

\reference{} Orosz, J., et al., 2001, ApJ, 555, 489

\reference{} Poutanen J., and Svensson, R., 1996, ApJ, 470, 249

\reference{} Revnivtsev, M., Gilfanov, M., Churazov, E., \& Sunyaev,
R., 2002, A \& A, subm., astro-ph/0204132

\reference{} Shrader, C., \& Titarchuk, L., 1999, apJ, 521, L21

\reference{} Smith, D. A., Levine, A. M., \& Morgan, E. H., 1999, IAU
Circ. 7253

\reference{} Swank, J., \& Markwardt, C., 2001, in the proceedings of
the conference ``New Century of X-ray Astronomy'', 2001 March 6--8,
Yokohama, Japan, astro-ph/0109240

\reference{} Titarchuk, L., 1994, ApJ, 434, 313

\reference{} Weaver, K., Gelbord, J., \& Yaqoob, T., 2001, ApJ, 550, 261

\reference{} In~'t Zand, J. J. M., et al., 2000, A \& A, 357, 526

\reference{} In~'t Zand, J. J. M., et al., 2002, A \& A, in press,
astro-ph/0205339

\end{references}
\end{document}